\documentstyle[seceq,supplement,epsf]{ptptex}

\title{
Absorbing Boundary Condition Approach for Breakup Reactions of
Halo Nuclei
}

\author{
Kazuhiro {\sc Yabana},
Manabu {\sc Ueda}
and Takashi {\sc Nakatsukasa}$^*$
}

\inst{
Institute of Physics, University of Tsukuba, Tsukuba 305-8571
\\
$^*$ Physics Department, Tohoku University, Sendai 980-8578
}

\recdate{
}

\abst{
Application of the absorbing boundary condition is discussed to analyse 
breakup reactions of weakly bound nuclei. The key ingredient is an 
introduction of the absorbing potential outside the physical area
which simulates the outgoing boundary condition approximately. 
The scattering problem is then recasted into the Schr\"odinger like 
equation with a source term in the interaction region and with the 
vanishing boundary condition at the boundary. We demonstrate 
usefulness of the method taking a few examples. Deuteron breakup 
reactions are examined comparing present results with those by the 
continuum-discretized coupled-channel method. We next discuss the 
breakup reactions of single-neutron halo nucleus, $^{11}$Be.}

\begin{document}

\maketitle

\section{Introduction}

Nuclei around drip line are characterized by their small separation
energy for breakup (fragmentation). Measurements of the breakup
processes have been major sources of information on their structural
properties.\cite{Tan95} Developments of reaction theories which are 
capable of describing breakup processes have thus been urged.

The eikonal approximation can take account of breakup processes and 
provides the most powerful scheme to describe reactions of weakly bound
projectile at high incident energies. In fact, the eikonal approximation
was employed to investigate the halo structure quantitatively through 
analyses of the anomalously large interaction and fragmentation cross 
sections\cite{BBS89,OYS92,YOS92,ATT96}.
As the precise measurements at lower incident energy have come to be 
available, demands for developments of reaction theories beyond 
eikonal approximation have been increasing. There have been several
attempts in this direction.\cite{Joh98,Esb99,Ber02} 
Among them, the coupled channel 
approach incorpolating breakup processes into continuum states, 
which is known as the continuum-discretized coupled-channel (CDCC) 
method\cite{CDCC}, is expected to be useful for this purpose. 
The CDCC method has been successful to describe direct reactions of 
nuclei with small separation energy for fragmentation, such as 
deuteron and $^{6,7}$Li.
The method has recently been applied to the reactions of halo nuclei, 
as is reported by Tostevin in this symposium\cite{Tos02,Tos01}. 
Although the CDCC approach has been well-tested and widely applied, 
it includes construction of artificial discretized continuum states 
and requires delicate examination for the convergence of the results.

In this report, we would like to show that the absorbing boundary
condition (ABC) approach, which was originally developed in the field 
of chemical reactions\cite{SM92}, provides a convenient and flexible 
descriptions of breakup reactions.
In this approach, we can directly obtain real-space scattering wave 
function in the interaction region without introducing any discretized 
continuum channels and scattering boundary condition. 
The trade-off for its conceptual simplicity is a heavy computational 
cost to calculate wave function in real-space. Since the problem can 
be recasted into a linear algebraic equation with a sparse, large-size 
matrix, we can make the best use of the state-of-the-art techniques 
to treat these problems.

The organization of this report is as follows. We first explain what
the ABC is, by taking a simple potential scattering problem as an example. 
We then apply it to breakup reactions of deuteron for which detailed 
CDCC analyses are available. We can examine validity 
and usefulness of our approach in this example. We then apply the method 
to reaction of single-neutron halo nuclei, $^{11}$Be, and discuss some 
characteristic features in the breakup reaction.

\section{Absorbing Boundary Condition: Potential Scattering}

We first explain a basic idea of the ABC for scattering problem, 
taking the simplest example, a point particle scattered by a potential 
$V({\bf r})$.

The wave function with outgoing boundary condition is expressed,
as usual, as a sum of the incident plane wave and the scattered outgoing 
wave as
\begin{equation}
\psi^{(+)}({\bf r}) = e^{ikz} + \psi_{\rm scat}({\bf r}).
\end{equation}
The scattered wave $\psi_{\rm scat}({\bf r})$ is expressed
employing the Green's function as
\begin{equation}
\psi_{\rm scat} = \frac{1}{E+i\epsilon-T} V \psi^{(+)}
= \frac{1}{E+i\epsilon-T-V} V e^{ikz},
\end{equation}
where $\epsilon$ is a positive infinitesimal and specifies the
outgoing boundary condition. 

The basic trick of the ABC is a replacement of the infinitesimal 
positive number $\epsilon$ with a finite, space-dependent function 
$\epsilon({\bf r})$. If the function is regarded as a part of the 
Hamiltonian, $T+V-i\epsilon({\bf r})$, the replacement is equivalent 
to adding absorbing potential $-i\epsilon({\bf r})$ to the Hamiltonian. 
To simulate outgoing boundary condition, one introduces sufficiently 
smooth and strong enough positive function $\epsilon({\bf r})$ outside 
a certain 
radius $R$ beyond which the potential $V({\bf r})$ can be ignored. 
Placing the absorbing potential in radial
region $R < r < R + \Delta R$, we put a vanishing boundary condition 
at $r=R+\Delta R$. If the absorbing potential works ideally, only 
outgoing wave can exist just inside the radius $R$. Since the scattering 
amplitude can be calculated from the exact wave function in the 
interaction region, it is sufficient that the ABC provides accurate 
wave functions in the spatial region $r<R$.

With the replacement $\epsilon \rightarrow \epsilon({\bf r})$, 
we can rewrite the equation for $\psi_{\rm scat}$ as the following 
linear inhomogeneous equation,
\begin{equation}
\left( E+i\epsilon({\bf r})-T-V \right) \psi_{\rm scat}
= V e^{ikz}.
\label{scateq}
\end{equation}
After the partial wave expansion and discretization in radial coordinate,
this is a linear algebraic problem with a complex symmetric coefficient 
matrix. We can then calculate scattered solution just by solving this 
linear algebraic problem.

The outgoing boundary condition imposed in this approach is not exact, 
since any imaginary potential can never absorb incoming waves completely.
One should choose the function $\epsilon({\bf r})$ so that the reflection
wave becomes as small as possible. The following linear absorbing potential
\begin{equation}
i\epsilon({\bf r}) = \left\{
\begin{array}{cl}
0 & (r < R) \\
i W_{abs} \frac{r-R}{\Delta R} & (R < r < R+\Delta R)
\end{array}\right.
\end{equation}
has been often used and well tested.\cite{Chi91,NY01} 
Here the absorbing potential works
in the region $R < r < R+\Delta R$. $W_{abs}$ is a positive constant and 
specifies 
the strength of the absorbing potential. In practical applications, the 
parameters should satisfy following conditions for a good 
absorber,\cite{Chi91,NY01}
\begin{equation}
20\frac{E^{1/2}}{\Delta R \sqrt{8m}} < W_{abs}
< \frac{1}{10}\Delta R \sqrt{8m}E^{3/2},
\label{Wcond}
\end{equation}
where $E$ represents the incident energy and $m$ is the relevant mass. 
The left inequality originates from the condition that the absorption is 
strong enough to suppress any reflection at $r=R+\Delta R$, while the 
right inequality from the condition that the reflection at $r=R$ is 
sufficiently small.
As the incident energy $E$ becomes lower, the wider absorbing area
$\Delta R$ is required to find appropriate value of $W_{abs}$ satisfying
condition (\ref{Wcond}), because of the increase of the wave length.

\begin{figure}[b]
\centerline{
\begin{minipage}{.45\linewidth}
\epsfxsize=60mm
\epsfbox{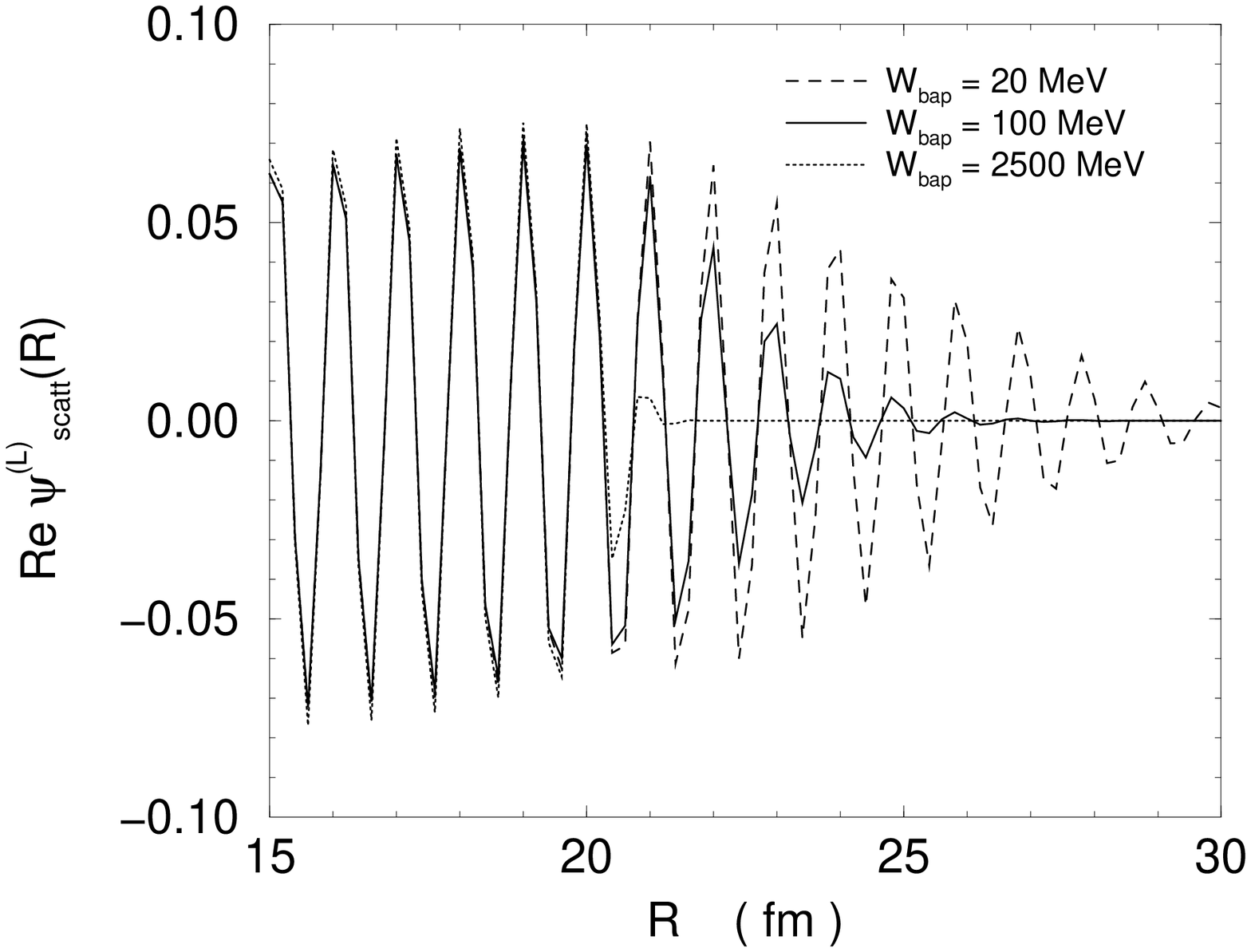}
\end{minipage}
\hspace{10mm}
\begin{minipage}{.45\linewidth}
\epsfxsize=60mm
\epsfbox{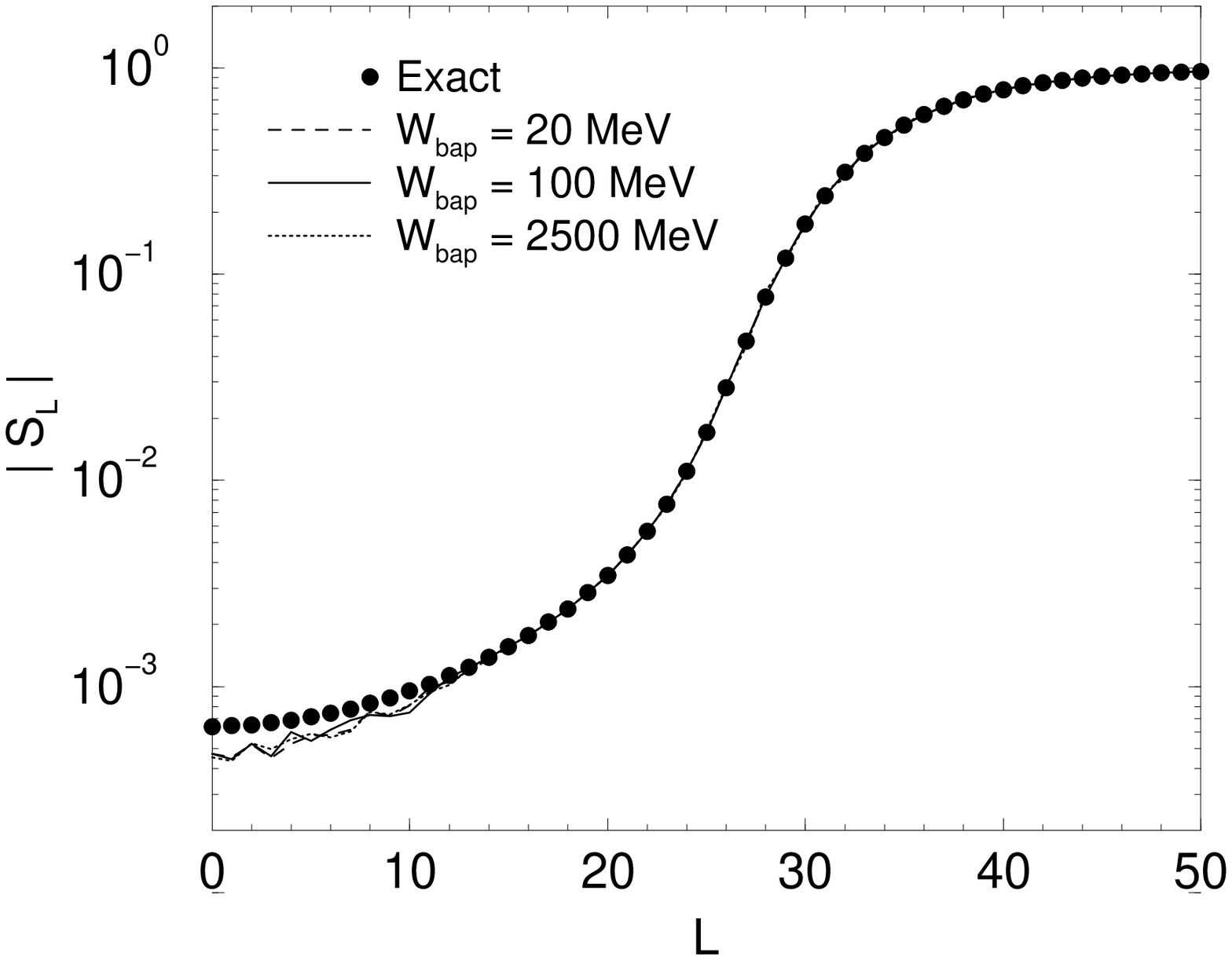}
\end{minipage}
}
\caption{
The potential scattering calculated with the ABC. 
$^{10}$Be-$^{12}$C scattering with the optical potential at incident 
energy $E=300$ MeV. [Left] The radial wave function for $L=20$ with 
three different choices of absorbing potential strength. 
[Right] Absolute value of scattering matrix calculated with the ABC
(curves), in comparison with accurate values (dots).
}
\label{fig:1}
\end{figure}

As a demonstration, we show in Fig.~\ref{fig:1} the scattering of
$^{10}$Be-$^{12}$C described with the optical potential. The radial 
wave function is shown for a partial wave of $L=20$ in the left hand panel,
and the absolute value of the scattering matrix in the right hand panel. 
The Woods-Saxon shape is used for the potential with parameters,
$V=123$ MeV, $r_V=0.75$fm, $a_V=0.80$fm, $W=65$MeV, $r_W=0.78$fm, 
and $a_W=0.80$fm. The incident energy is set at $E_{\rm lab}=300$ MeV.
In solving Equation~(\ref{scateq}) in the partial wave expansion, 
the plane wave $e^{ikz}$ in the right hand side is replaced by the 
regular Coulomb wave function and $V$ in the right hand side represents 
the potential in which the point Coulomb potential is subtracted. 
The radial equation is discretized with $\Delta r = 0.2$ fm grid, 
and the discrete variables representation is used for kinetic energy
operator.\cite{SM92} We put an absorber outside $R=20$ fm with 
$\Delta R=10$ fm thick. Results with several choices for $W_{abs}$ are 
compared in the figure, $W_{abs}=20, 100, 2500$ MeV. 
The choice of $W_{abs}=100$ MeV satisfies the criterion given by 
Eq.~(\ref{Wcond}), while others do not. As seen from the right hand 
panel of Fig.~\ref{fig:1}, the scattering matrix is rather insensitive 
to the choice of the absorbing potential. One may thus obtain 
accurate scattering matrix with a rather wide window of $W_{abs}$.

\section{Absorbing Boundary Condition: Breakup Reactions}

The advantage of the ABC manifests itself in the three-body scattering
problems. Let us consider reaction of a projectile (P) composed of 
core (C) plus neutron (n), P=C+n, on a target nucleus (T). 
Expressing projectile-target relative coordinate by ${\bf R}$ and 
neutron-core relative coordinate by ${\bf r}$, the Hamiltonian 
of this three-body system is expressed as
\begin{equation}
H = -\frac{\hbar^2}{2\mu}\nabla_{\bf R}^2
      -\frac{\hbar^2}{2m} \nabla_{\bf r}^2
      +V_{nC}({\bf r}) + V_{nT}({\bf r}_{nT}) 
      +V_{CT}({\bf R}_{CT})
\end{equation}
where $\mu$ and $m$ are reduced masses of projectile-target relative 
motion and neutron-core relative motion, respectively. $V_{nC}$,
$V_{nT}$, $V_{CT}$ are the interaction potentials of constituent
particles. $V_{CT}$ includes the Coulomb potential. $V_{nT}$ and $V_{CT}$
may include imaginary potentials which represent loss of flux from the
model space, while $V_{nC}$ is taken to be real.

We introduce redundant 'distorted' wave with outgoing boundary condition
by the following equation,
\begin{equation}
\left\{ -\frac{\hbar^2}{2\mu} \nabla_R^2 + V_{PT}({\bf R}) \right\}
\psi^{(+)}({\bf R}) = E \psi^{(+)}({\bf R}),
\end{equation}
where $V_{PT}$ may be chosen arbitrary but should include the Coulomb 
potential. The total wave function may be expressed as a 
sum of the distorted wave in the incident channel and the scattered wave.
\begin{equation}
\Psi^{(+)}({\bf R},{\bf r})
= \psi^{(+)}({\bf R}) \phi_0({\bf r}) + \Psi_{\rm scat}({\bf R},{\bf r})
\end{equation}
where $\phi_0({\bf r})$ is the initial bound orbital of neutron-core
relative motion. The scattered wave $\Psi_{\rm scat}$ satisfies the
following inhomogeneous equation in the ABC,
\begin{eqnarray}
&& \left\{ E + e_0 + i\epsilon_{nC}(r) + i\epsilon_{PT}(R) -H \right\}
\Psi_{\rm scat}({\bf R},{\bf r})  \nonumber\\
&&=
\left\{ V_{nT}({\bf r}_{nT})+V_{CT}({\bf R}_{CT}) 
- V_{PT} \right\}
\psi^{(+)}({\bf R}) \phi_0({\bf r}),
\label{3Bscat}
\end{eqnarray}
where $E$ is the energy of projectile-target relative motion
in the incident channel and $e_0$ is the binding energy in the
projectile. One should note that the right hand side
$\left\{ V_{nT} + V_{CT} - V_{PT} \right\} 
\psi^{(+)}({\bf R}) \phi_0({\bf r})$
is a localized function in space. Namely, this function vanishes
if either $R$ or $r$ is large enough. The absorbing potentials,
$\epsilon_{nC}+\epsilon_{PT}$, assure the outgoing boundary condition
to be satisfied approximately at spatial region where either $R$ or $r$ 
is large.

In practice, Equation (\ref{3Bscat}) is solved in the partial wave
expansion, expressing the wave function $\Psi^{(+)}({\bf R},{\bf r})$ as
\begin{equation}
\Psi^{(+)}({\bf R},{\bf r})
= \sum_{Ll} \frac{u^J_{Ll}(R,r)}{Rr}
\left[ Y_L(\hat R) Y_l(\hat r) \right]_J.
\label{pwe}
\end{equation}
Denoting the angular momentum channels specified by $L$ and $l$ as
$a$, and assuming $s$-wave state for $\phi_0({\bf r})$,
the equation for $u^J_{Ll}(R,r)$ reads
\begin{eqnarray}
&& \Bigl\{ E +e_0 + i\epsilon_{nC}(r) + i\epsilon_{PT}(R)
\nonumber\\
&&  -\left( 
-\frac{\hbar^2}{2\mu}\frac{\partial^2}{\partial R^2}
+ \frac{\hbar^2 L_a(L_a+1)}{2\mu R^2} 
-\frac{\hbar^2}{2m} \frac{\partial^2}{\partial r^2}
+ \frac{\hbar^2 l_a(l_a+1)}{2m r^2} + V_{nC}(r) \right) \Bigr\}
u^J_a(R,r)
\nonumber\\
&&  -\sum_{a'} V^J_{aa'}(R,r)
u^J_{a'}(R,r) \nonumber\\
&&=
\left\{ V^J_{aa_0}(R,r) - \delta_{a a_0} V_{PT}(R) \right\} w_J(R) v_0(r)
\label{3Brad}
\end{eqnarray}
where $a_0$ is the angular momentum channel of incident wave,
$v_0$ is the radial wave function of $\phi_0({\bf r})$,
$w_J(R)$ is the radial wave function of $\psi^{(+)}({\bf R})$
for partial wave $J$. $V^J_{aa'}(R,r)$ is a multipole expansion of
the potential $V_{nT}+V_{CT}$.

The boundary condition is given as, for arbitrary $R$ and $r$, 
$u^J_{Ll}(R,0)=u^J_{Ll}(0,r)=0$ at origin and 
$u^J_{Ll}(R,r_{max})=u^J_{Ll}(R_{max},r)=0$ at the
boundaries, $R_{max}$ and $r_{max}$.

\section{Deuteron Reaction}

In this section we apply the ABC approach to breakup reactions of deuteron. 
Since the deuteron reaction has been investigated in detail with the 
CDCC method,\cite{CDCC,Yah82} we can assess reliability of the ABC by 
comparing results of both methods.

We consider $d$+$^{58}$Ni reaction at incident deuteron energy $E_d = 80$ 
MeV. The wave function is expanded in partial waves, as given in 
Eq.~(\ref{pwe}). The spatial parameters of the wave function are set as 
follows: The relative angular momenta between proton and neutron of 
$l$=0 and 2 are included. The radial wave function $u^J_{Ll}(R,r)$ is 
discretized with grid spacing of 0.2 fm for $R$ and 0.5 fm for $r$. 
The radial region up to 50 fm are treated, and each absorbing potential
is placed in the regions larger than 25 fm. The absorbing potentials
are thus characterized by $\Delta r = \Delta R = 25$ fm. Their strengths of
$W_R = 50$ MeV and $W_r = 20$ MeV are employed. The potential 
parameters of $p,n$-$^{58}$Ni are taken to be the same as those adopted
in Ref.~\citen{Yah82}. The proton-neutron potential is a central force, 
and $s$-wave is assumed in the ground state.

We first discuss computational aspects of our method. Denote the 
number of radial grid points for $R$ as $N_R$ and that for $r$ as $N_r$, 
and the number of angular momentum channels specified by $Ll$ as $N_J$.
The wave function is then expressed as a column vector of dimension 
$N_{\rm dim}=N_R N_r N_J$. The Hamiltonian operator is a sparse, complex 
symmetric matrix. The scattering problem of Eq.(\ref{3Brad}) thus 
results in the linear algebraic problem of this dimension. 
For the present deuteron reaction, $N_R=250$, $N_r=100$ and $N_J=4$
give $N_{\rm dim}=100,000$. For such problem of large size, iterative methods 
are useful. We employ Bi-CG (Conjugate Gradient) method which is useful 
for problems with complex non-hermite coefficient matrix. 
A preconditioning utilizing diagonal elements of the coefficient matrix 
makes the convergence faster, since it gets rid of huge diagonal
elements at small $R$ and $r$ region due to the centrifugal barrier.

\begin{figure}[t]
\epsfxsize=110mm
\centerline{
\epsfbox{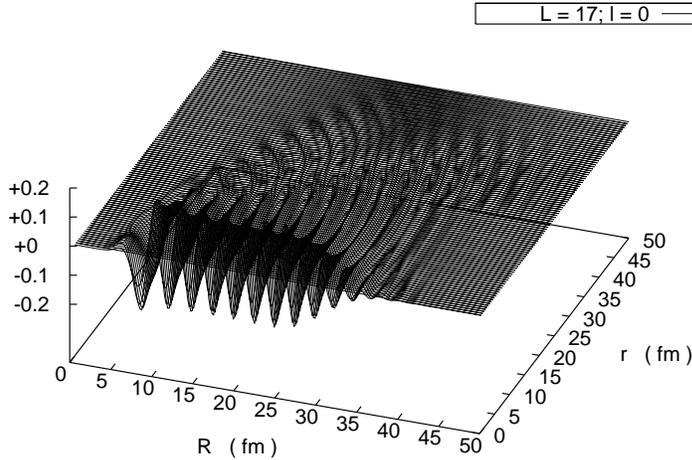}
}
\caption{
The real part of the radial wave function $u^J_{Ll}(R,r)$ for $d$-$^{58}$Ni 
scattering at $E_d=80$ MeV for $J=17$, $L=17$, and $l=0$.
}
\label{fig:2}
\end{figure}

We show in Fig.~\ref{fig:2} the wave function in the partial wave expansion,
${\rm Re}[u_{Ll}^J(R,r)]$ for $J=17$, $L=17$, and $l=0$. This includes 
both the incident and breakup waves. In the small $r$ region, the 
wave function is dominated by the incident wave, and shows oscillation 
as a function of $R$ reflecting the incident relative wave function. 
The amplitude of the wave function decreases at large 
$R$ due to the absorption by $\epsilon_{PT}(R)$. The wave function 
at large $r$ shows breakup components of deuteron into $p+n$ continuum 
state. The amplitude of the wave function also decreases at large $r$ 
due to the absorption by $\epsilon_{nC}(r)$.

\begin{figure}
\centerline{
\begin{minipage}{.45\linewidth}
\epsfxsize=60mm
\epsfbox{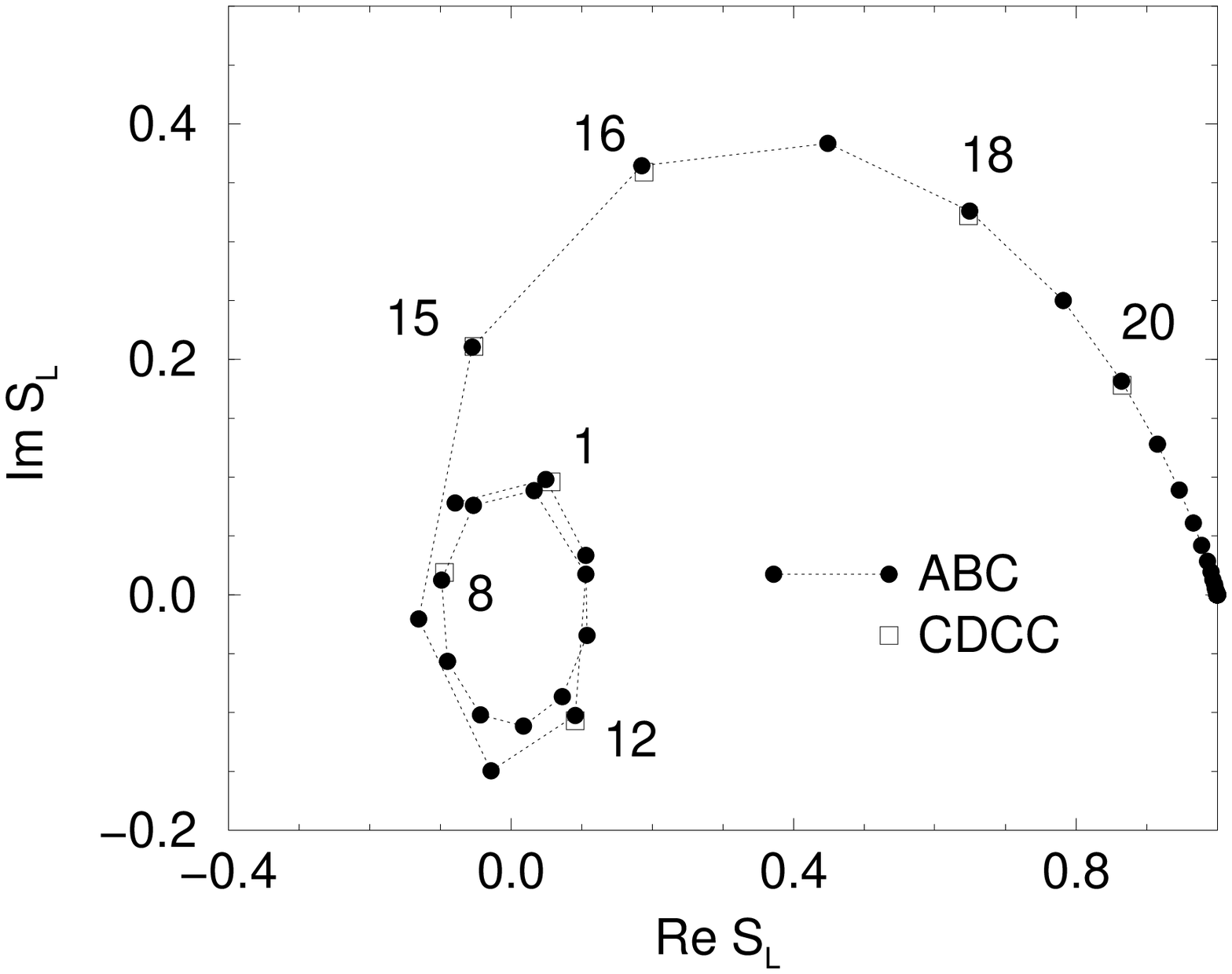}
\end{minipage}
\hspace{10mm}
\begin{minipage}{.45\linewidth}
\epsfxsize=60mm
\epsfbox{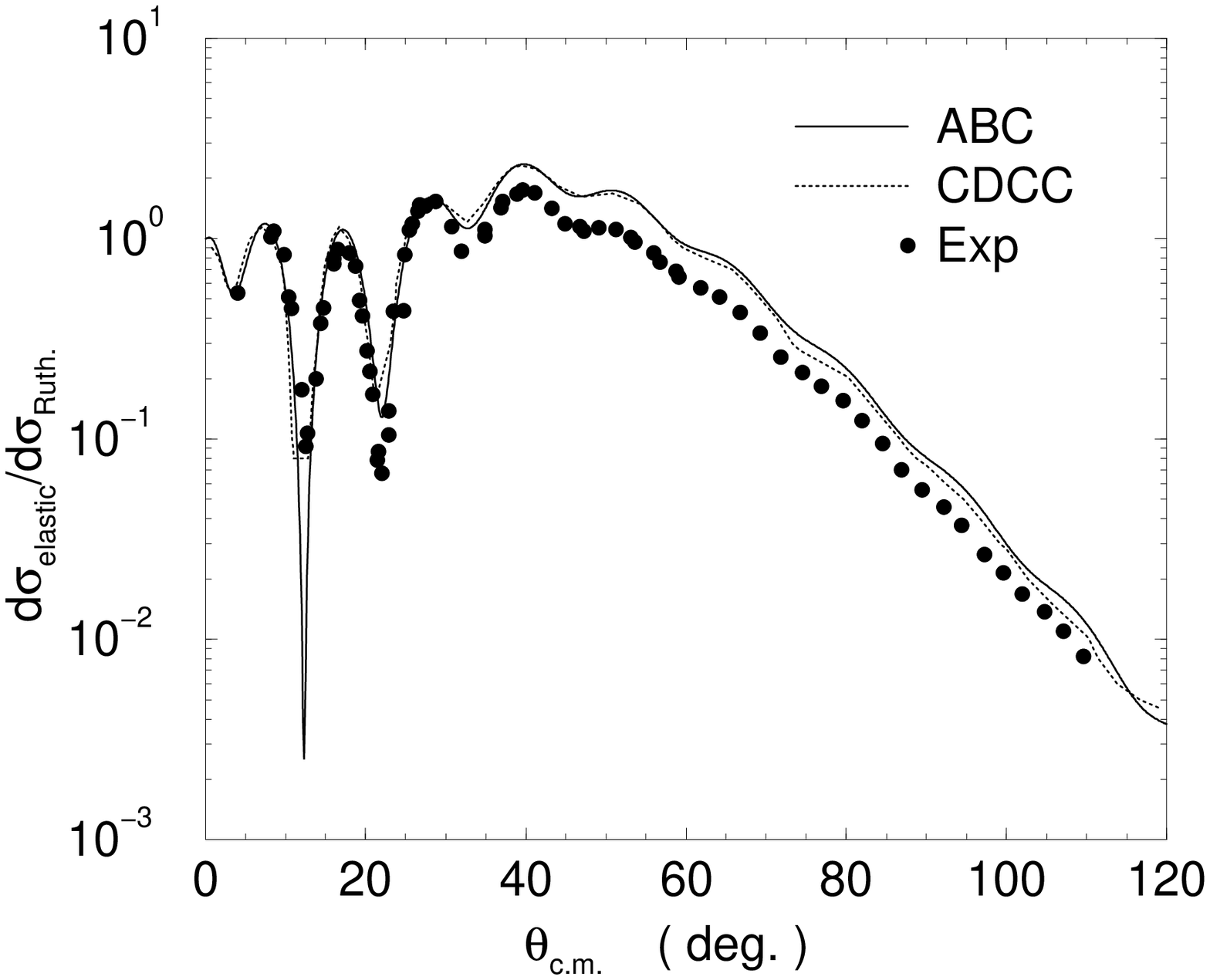}
\end{minipage}
}
\caption{
The $S$-matrix (left) and differential cross section (right) of
$d$-$^{58}$Ni elastic scattering at $E_d=80$ MeV. Calculations with
the ABC are compared with those by the CDCC\cite{Yah82}. Measured 
differential cross section is also plotted.
}
\label{fig:3}
\end{figure}

In Fig.~\ref{fig:3}, we show comparison of elastic scattering S-matrix
and cross section between the CDCC\cite{Yah82} and the ABC calculations.
These figures shows that the ABC calculation accurately reproduces
the CDCC results. We thus confirm that the ABC approach has the same
physical contents as those of the CDCC method, although the discretized
continuum channels are never introduced in the ABC calculation.
We also calculated the scattering matrices into deuteron breakup channels
specified by the relative momentum between proton and neutron.
We again confirmed that the ABC gives the same scattering matrices as
those by the CDCC.

\section{Application to $^{11}$Be + $^{12}$C reaction}

We here report analysis of reactions of single-neutron halo nucleus 
$^{11}$Be on $^{12}$C target. The $s$-wave bound state of $^{10}$Be + n 
relative motion is assumed for $^{11}$Be ground state. The $^{10}$Be-n
potential is taken as Woods-Saxon shape whose depth is set to give
the $2s$ orbital binding energy at 0.504 MeV. Optical potential
for $^{10}$Be -$^{12}$C is the same as that in Sect.2 (without energy
dependence), and the Bechetti-Greenlees potential is used for $n-^{12}$C.
The radial grid is chosen as 0.2 fm for $R$ and 0.5 fm for $r$.
The radial region up to 30 fm and 50 fm are used for $R$ and $r$, 
respectively. The absorbing potential is placed in the region 
20 fm $< R <$ 30 fm and 25 fm $< r <$ 50 fm.
The neutron-$^{12}$C relative angular momenta up to $l=3$ are included.
The matrix size for the wave function is thus about 240,000.

\begin{figure}[t]
\epsfxsize=110mm
\centerline{
\epsfbox{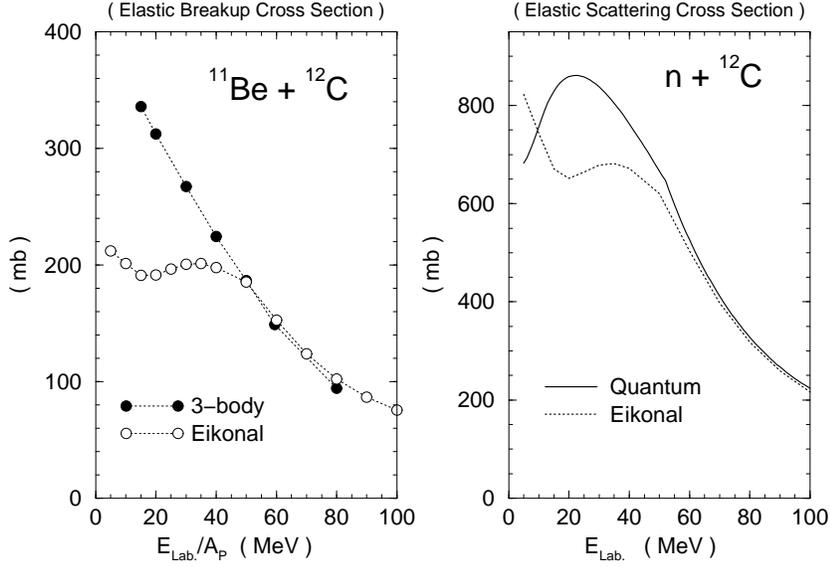}
}
\caption{
[Left] The elastic breakup cross section in $^{11}$Be-$^{12}$C reaction.
[Right] The neutron-$^{12}$C elastic scattering cross section. Quantum 
calculation is compared with eikonal calculation. 
}
\label{fig:4}
\end{figure}

The total reaction and the elastic breakup cross sections are expected
to be well described with the eikonal approximation at medium and
high incident energies. Below a certain incident energy, treatment
beyond the eikonal approximation would be required. One may expect
that the validity of the eikonal approximation at low incident energy
can be examined by investigating separately the two-body scatterings of
constituent particles in the eikonal approximation. 
In the right-hand panel of Fig.~\ref{fig:4}, the elastic scattering 
cross section of neutron-$^{12}$C scattering is shown. This cross section 
is expected to be related to the elastic breakup cross section of 
$^{11}$Be into $^{10}$Be+n fragments, because the elastic breakup 
process is considered to be dominated by the elastic scattering of 
halo neutron by the target nucleus. The eikonal approximation comes
to lose accuracy below incident energy of 50 MeV and underestimates
the cross section. The energy dependence of the cross section also looks
different between the exact and eikonal calculations. The eikonal 
approximation is expected to be more accurate for $^{10}$Be-$^{12}$C 
reaction cross section because of the shorter wave length at the same 
incident energy per nucleon for this system.

We now move to the three-body reaction problem. In the left-hand
panel of Fig.~\ref{fig:4}, we show the elastic breakup cross sections 
of $^{11}$Be-$^{12}$C reaction. The filled circles are the calculation
with the ABC and the open circles by the eikonal approximation.
The elastic breakup cross section is substantially larger than that in the 
eikonal approximation at lower incident energy. This clearly shows the
necessity to solve the problem beyond eikonal approximation below the
incident energy less than 50 MeV per nucleon. Comparing left and right
panels, qualitative features in the cross sections are similar between
the elastic breakup cross section of $^{11}$Be-$^{12}$C and the 
neutron-$^{12}$C elastic cross sections. This confirms the above argument
that the elastic breakup process is dominated by the elastic scattering
of neutron-target. However, looking at quantitatively, the discrepancy
between the exact and eikonal approximation is much larger for the
neutron breakup reactions. We think it necessary to examine further
the convergence of the calculations with respect to, for example, the 
relative angular momentum of neutron-core.

\section{Summary}

We have presented the ABC approach to describe breakup reactions
of weakly bound projectile. With the ABC, a general scattering 
problem with outgoing boundary condition is recasted into an 
linear inhomogeneous differential equation with vanishing boundary 
condition. Thus the scattering problem may be treated without 
introducing any asymptotic form for the wave function. 
In the partial wave expansion and radial discretization, the linear
inhomogeneous equation results in the linear algebraic problem with
sparse, complex symmetric coefficient matrix. Efficient iterative 
solvers are useful to solve this linear problem.

We show application of our method to deuteron reaction for which
detailed analyses with the CDCC method is available. We have confirmed
that the ABC calculation accurately reproduce the CDCC results.
Thus the ABC approach provides an alternative approach to describe 
reactions where breakup processes into continuum channels are important.
Comparing two approaches, the ABC is advantageous in that the
scattering wave function is obtained directly in real-space, avoiding
introduction of virtual continuum channels. The convergence of the
calculation can be examined with a few, intuitive, parameters related
to the shape of absorbing potential. The trade-off for its simplicity
is a heavy computational task to solve the large linear problem.

We have applied the ABC approach for breakup reaction of 
single neutron-halo nucleus $^{11}$Be on $^{12}$C target. 
It is found that the eikonal approximation becomes inaccurate 
at incident energy below 50 MeV/A. The eikonal approximation 
substantially underestimates the breakup cross section. 

Although we here discuss only reaction problems, the ABC will be useful
for any circumstances where the scattering boundary condition comes
into play. Responses in the continuum is one of the other problems 
where the ABC is extremely useful, as discussed in another report 
in this symposium.\cite{N02,NY01}

\end{document}